# *In situ* accurate determination of the zero time delay between two independent ultrashort laser pulses by observing the oscillation of an atomic excited wave packet

**Qun Zhang[1,*] and John W. Hepburn[2]**

[1]*Hefei National Laboratory for Physical Sciences at the Microscale and Department of Chemical Physics, University of Science and Technology of China, Hefei, Anhui 230026, P. R. China*

[2]*Departments of Chemistry and Physics, The University of British Columbia, Vancouver, B.C. V6T 1Z1, Canada*

*\*Corresponding author: qunzh@ustc.edu.cn*

We propose a novel method that uses the oscillation of an atomic excited wave packet observed through a pump-probe technique to accurately determine the zero time delay between a pair of ultrashort laser pulses.   This physically based approach provides an easy fix for the intractable problem of synchronizing two different femtosecond laser pulses in a practical experimental environment, especially where an *in situ* time zero measurement with high accuracy is required.

Much attention in the field of "quantum control" has recently been drawn to merging two existing successful strategies, i.e., "coherent control" [1] and "adiabatic passage" [2], and utilizing the resulting "coherently controlled adiabatic passage" method to achieve both selectivity and completeness [3,4]. Following this theoretical framework, a control scenario aimed at achieving near unity efficiency of population transfer has very recently been proposed [5] and implemented [6]. Such types of experiments and alike can be viewed as extending the well-established concept of stimulated Raman adiabatic passage (STIRAP) [7,8] from the nanosecond (or picosecond) domain to the femtosecond (fs) domain, which necessitates an accurate measurement of timing between a (pump-probe) pair of (independent) fs laser pulses.

In an ongoing experiment with a similar goal to that reported in [6], we apply the conventional pump-probe technique to a ladder system (3s→3p→3d) of atomic sodium ($^{23}$Na) in an attempt to achieve complete population transfer from the initially populated 3s state to the initially unpopulated target 3d state. For such an fs-STIRAP type experiment, accurate determination of the zero time delay (ZTD) between the two fs pulses plays a critical role in the search of the desirable "counterintuitive" enhancement indicating a complete population transfer, a signature of STIRAP [7, 8]. Such a signature is expected to lie around a certain negative time delay (i.e., the Stokes or probe pulse precedes the pump pulse) in the close vicinity (on the fs pulse duration time scale) of the time zero point.

To the best of our knowledge, when conducting fs pump-probe experiments one determines the ZTD usually by means of the well-known sum-frequency mixing cross-correlation (SFM-CC) technique [9] that uses a thin (or ultra-thin) nonlinear crystal (e.g., BBO). Although this facile approach is rather straightforward and does work for most of pump-probe applications, it is proved practically invalid in the aforementioned fs-STIRAP experiments, where a limited space of laser-atomic beam interaction region inside a vacuum chamber precludes one from conveniently installing a thin nonlinear crystal and efficiently collecting the very weak SFM-CC signal.   It is also worth noting that, an alternative way to measure the SFM-CC signal in front of the laser entrance window of the vacuum chamber can hardly be unambiguous for determining the desirable ZTD, simply because it neglects the different amounts of dispersion introduced by the window material to the pump and probe pulses of different wavelengths.

In this Letter, we propose and demonstrate a physically based method, which takes advantage of the oscillation of an atomic excited wave packet observed through the pump-probe technique to accurately determine the actual ZTD between a pair of ultrashort laser pulses used in the pump-probe experiments *in situ*.

The left part of Fig. 1 depicts the pump-probe scheme on the ladder system ($3s \rightarrow 3p \rightarrow 3d$) of atomic sodium ($^{23}$Na).   The pump pulse, which comes from a Ti:sapphire-pumped optical parametric amplifier (TOPAS, Light Conversion, 1kHz) operating at a center wavelength of ~593 nm with an FWHM duration of ~130 fs and a peak intensity of ~$2 \times 10^{10}$ W/cm$^2$, couples the ground 3s state with the two intermediate fine structure $3p_{1/2}$ and $3p_{3/2}$ levels.   As a result, a wave packet due to the coherent superposition of the excited $3p_{1/2}$ and $3p_{3/2}$ levels is prepared. The Stokes pulse, which comes from a Ti:sapphire regenerative amplifier (Spitfire, Specra-Physics, 1kHz) operating at a center wavelength of ~812 nm with an FWHM duration of ~150 fs and a peak intensity of ~$3 \times 10^{12}$ W/cm$^2$, couples the two intermediate levels with the target 3d state, and the same pulse simultaneously works as a probe to ionize the system to a set of orthogonal continuum states. A time delay ($\tau$) between the two pulses was introduced by using a translating retroreflector placed in the pump laser beam path. The two laser beams were then recombined and focused onto the Na atomic beam generated from an oven heated to ~600 K. When the two pulses are temporally well separated, the total measured Na ion signal is expected to oscillate as a function of the time delay $\tau$. The oscillations originate from quantum beats between two paths sharing the same initial and final states, and involving the fine structure $3p_{1/2}$ and $3p_{3/2}$ levels respectively.

Although such a stationary state treatment can well describe the observed quantum beats (see Fig. 2), it is incapable of deriving phase (and amplitude) information from the oscillations. Alternatively, a time dependent description in the bright-dark states basis developed by Zamith et al. [10] has proved to be able to provide a more direct physical insight into the interaction process described above. In the present case of two excited states ($3p_{1/2}$ and $3p_{3/2}$), the bright and dark states can be defined respectively by $|\psi_B\rangle = \cos\alpha |3p_{1/2}\rangle + \sin\alpha |3p_{3/2}\rangle$ and $|\psi_D\rangle = -\sin\alpha |3p_{1/2}\rangle + \cos\alpha |3p_{3/2}\rangle$, where $\tan\alpha = \mu_{3p_{3/2},3s} / \mu_{3p_{1/2},3s}$ with $\mu_{3p_{3/2},3s}$ and $\mu_{3p_{1/2},3s}$ representing respectively the matrix element for the $|3s\rangle \rightarrow |3p_{3/2}\rangle$ and $|3s\rangle \rightarrow |3p_{1/2}\rangle$ transition of the dipole moment operator $\mu$ [10]. The pump laser field couples only the ground 3s state and the bright state, while the atomic spin-orbit Hamiltonian couples the bright and dark states, as depicted in the right part of Fig. 1.

Considering that the spectral width of the pump pulse (~280 cm$^{-1}$) used in this Letter is much larger than the energy separation of the two excited 3p substates ($\Delta E$ ~17.196 cm$^{-1}$ [11]), we can separate the laser interaction from the free evolution of the system. The interaction of the system with the pump pulse results in the creation of a wave packet localized in the bright state. As a result of the coupling between the bright and dark states, the wave packet evolves freely, oscillating periodically in and out of the bright state [12]. Such oscillations can be readily detected by the probe (Stokes) pulse centered at $\tau$ via a (1 + 1) ionization process, as shown in Fig. 1.

Following the bright-dark states formalism given in [10], we can determine that the total population in the ionization continuum can be expressed as $P(\tau) = P_B + [(P_D - P_B)\sin^2 2\alpha - P_{BD}\sin 4\alpha](1-\cos\Delta E\tau)/2$, where $P_B$ and $P_D$ represents respectively the total ionization probability from the bright and dark states, $P_{BD}$ the interference term corresponding to the possibility of accessing the same final state simultaneously from the bright and dark states. Notably, depending on the sign of the quantity $[(P_D - P_B)\sin^2 2\alpha - P_{BD}\sin 4\alpha]$, the oscillations with a beat frequency of $\Delta E$ must have either a maximum or a minimum at $\tau = 0$. It is this derivative conclusion that we can use for accurately determining the desirable actual ZTD between two ultrashort laser pulses.

The oscillations shown in Fig. 2 can be fitted with a cosine function, giving rise to a period of 1.94 ps, which can be Fourier-transformed to yield a beat frequency of 17.2 cm$^{-1}$, in excellent agreement with the 17.196 cm$^{-1}$ [11] value for the energy splitting of the Na 3p doublet. The ZTD to be determined is nominally set with respect to an SFM-CC measurement carried out beforehand outside the vacuum chamber, see also the insert of Fig. 3. To derive the actual ZTD, we extrapolate the fitted data (i.e., the solid cosine curve in Fig. 2) by two more cycles of

- 5 -

oscillation to the smaller time delay end, as indicated by the dotted cosine curve in Fig. 2. Based on the time dependent bright-dark states analysis given above and the fact that in the limit of ultrashort pulses, one should observe a sharp increase of the ion signal at $\tau = 0$, followed by a free evolution in the form of a cosine function [10], we can readily narrow down the search of the actual ZTD to only the two maxima (labeled a1 and a2 in Fig. 2) and the one minimum (labeled b) in between. The two maxima a1 and a2 lie far away (about −2 040 and +1 840 fs, respectively) from the nominal ZTD, whilst the minimum b lies at a reasonably close delay time of −98 fs (within the pulse duration), as can be clearly seen in Fig. 3 which is the expanded view of the time delay range between −400 and 200 fs in Fig. 2.

Although the actual ZTD at the laser-atomic beam interaction region cannot be determined by the *ex situ* SFM-CC measurement, it should not deviate too much from such a measurement. An estimate based on taking into account the dispersion introduced by the window material indicates that the discrepancy between the actual and nominal ZTD values lies within the pulse duration, i.e., smaller than 150 fs in the present case. Therefore, with the aid of the SFM-CC measurement, we can unambiguously identify the time delay value that corresponds to the minimum b in Fig. 2 as the actual time zero (marked by the vertical solid line in Figs. 2 and 3).

Since the ZTD measurement reported here is explicitly transformed to measuring the quantum beats extrema, its accuracy is equivalent to that of the phase measurement of the cosine function fitted to the experimental data. The latter is found to be approximately ±3% in the present study. Achieving such high accuracy is by no means a trivial task for other non-physically based techniques. More importantly, thanks to its *in situ* nature this approach gets rid of any ambiguities, thereby providing a stringent reference point for fs pump-probe applications, such as our ongoing fs-STIRAP experiment.

To summarize, we have presented a unique *in situ* determination of the desirable actual zero time delay between a pair of independent femtosecond laser pulses. Because this determination is achieved by observing the oscillation of an atomic excited wave packet (with the aid of a sum-frequency mixing cross-correlation measurement), high accuracy is guaranteed. We expect that this physically-based method will be widely applicable to many other studies in need of solving the vital problem of synchronizing two different femtosecond laser pulses in a practical experimental environment.

This research was made possible by financial support from the Canada Foundation for Innovation. The authors have enjoyed inspiring discussions with M. Shapiro. Q. Zhang thanks the University of Science and Technology of China (USTC) for awarding a start-up fund for the returned overseas Chinese scholars and a USTC Youth Fund. Experimental assistance from M. Keil, C. Li, S. Zhdanovich, V. Milner, and Q. J. Hu is also gratefully acknowledged.

## References

1. M. Shapiro and P. Brumer, *Principles of the Quantum Control of Molecular Processes* (Wiley, 2003).

2. N. V. Vitanov, T. Halfmann, B. W. Shore, and K. Bergmann, Annu. Rev. Phys. Chem. **52**, 763 (2001).

3. M. Shapiro and P. Brumer, Phys. Rep. **425**, 195 (2006).

4. P. Kral, I. Thanopulos, and M. Shapiro, Rev. Mod. Phys. **79**, 53 (2007).

5.  E. A. Shapiro, V. Milner, C. Menzel-Jones, and M. Shapiro, Phys. Rev. Lett. **99**, 033002 (2007).

6.  S. Zhdanovich, E. A. Shapiro, M. Shapiro, J. W. Hepburn, and V. Milner, Phys. Rev. Lett. **100**, 103004 (2007).

7.  K. Bergmann, H. Theuer, and B. W. Shore, Rev. Mod. Phys. **70**, 1003 (1998).

8.  N. V. Vitanov, M. Fleischhauer, B. W. Shore, and K. Bergmann, Adv. At. Mol. Opt. Phys. **46**, 55 (2001).

9.  J.-C. Diels and W. Rudolph, *Ultrashort Laser Pulse Phenomena: Fundamentals, Techniques, and Applications on a Femtosecond Time Scale* (Academic, 1996).

10. S. Zamith, M. A. Bouchene, E. Sokell, C. Nicole, V. Blanchet, and B. Girard, Eur. Phys. J. D **12**, 255 (2000).

11. NIST Atomic Spectra Database (NIST, Gaithersburg, Md., 2008).

12. C. Nicole, M. A. Bouchene, S. Zamith, N. Melikechi, and B. Girard, Phys. Rev. A **60**, R1755 (1999).

**Figure captions**

**Fig. 1.**  Pump-probe scheme on the 3s→3p→3d ladder system of atomic sodium.    The left part represents the stationary state description, while the right part represents the bright-dark states description.    See discussions in the text.

**Fig. 2.**  (Color online).    Pump-probe results in the sodium 3p manifold (circles, experimental data; solid curve, cosine function fit; dashed curve, extrapolated from the fitted curve).    The quantum beats period (~1.94 ps) derived from the fit corresponds to the fine structure splitting (~17.2 cm$^{-1}$) of the 3p state.    The positive time delays correspond to the pump pulse (centered at ~593 nm) preceding the Stokes (probe) pulse (centered at ~812 nm).    The $\tau = 0$ on the abscissa is a nominal value derived from an *ex situ* SFM-CC measurement.    The actual time zero determined in this Letter is indicated by the vertical solid line.

**Fig. 3.**  (Color online).    Expanded view of the time delay range [–400, 200 fs] in Fig. 2.    The actual time zero (corresponding to a nominal  $\tau = -98$  fs ) is indicated by the vertical solid line. The insert shows the result of an *ex situ* SFM-CC measurement, which yields the nominal  $\tau = 0$ given on the abscissa.

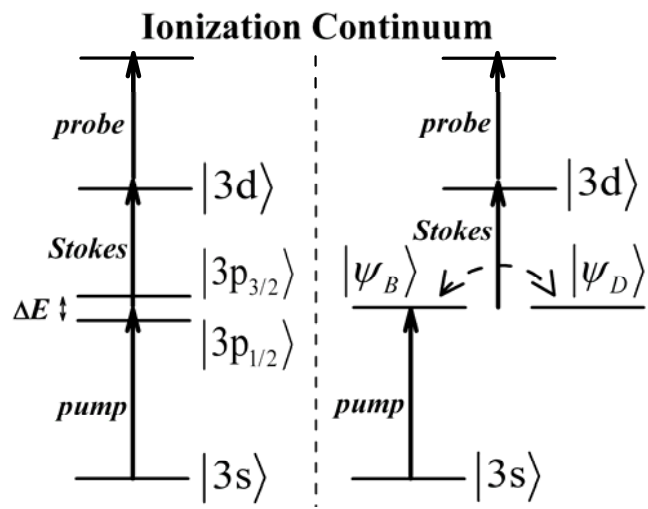

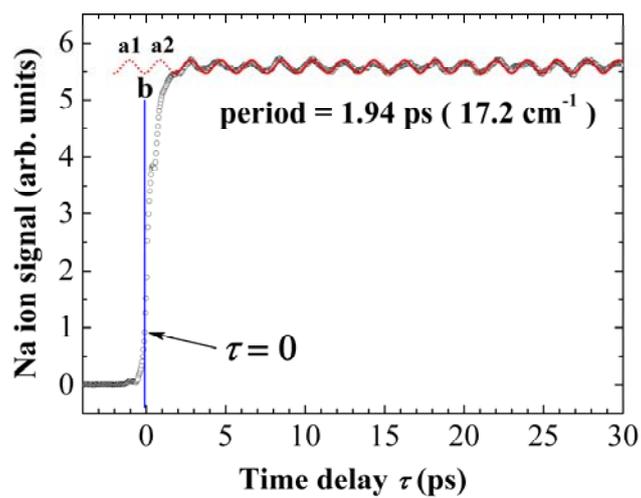

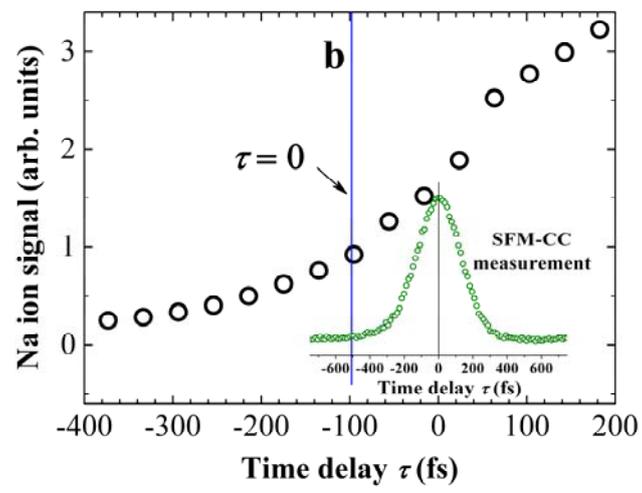